\documentclass[sigconf]{acmart}

\usepackage{color}
\AtBeginDocument{%
  }

\copyrightyear{2025}
\acmYear{2025}
\setcopyright{acmlicensed}\acmConference[VRST '25]{31st ACM Symposium on
Virtual Reality Software and Technology}{November 12--14, 2025}{Montreal,
QC, Canada}
\acmBooktitle{31st ACM Symposium on Virtual Reality Software and Technology
(VRST '25), November 12--14, 2025, Montreal, QC, Canada}
\acmDOI{10.1145/3756884.3766010}
\acmISBN{979-8-4007-2118-2/2025/11}

\begin{document}

\title{Unmanned Aerial Vehicles Control in a Digital Twin: Exploring the Effect of Different Points of View on User Experience in Virtual Reality}

\author{Francesco Vona}
\orcid{0000-0003-4558-4989}
\affiliation{%
  \institution{Immersive Reality Lab\\ University of Applied Sciences Hamm-Lippstadt}
  \city{Hamm}
  \country{Germany}
}
\email{francesco.vona@hshl.de}

\author{Mohamed Amer}
\orcid{0009-0007-3729-0354}
\affiliation{%
  \institution{University of Applied Sciences Hamm-Lippstadt}
  \city{Hamm}
  \country{Germany}
}
\email{mohamed-ahmed-mohamed-ali.amer@stud.hshl.de}

\author{Omar Abdellatif}
\orcid{0009-0004-3780-4886}
\affiliation{%
  \institution{University of Applied Sciences Hamm-Lippstadt}
  \city{Hamm}
  \country{Germany}
}
\email{omar.abdellatif@stud.hshl.de}

\author{Michelle Celina Hallmann}
\orcid{0009-0002-9600-7575}
\affiliation{%
  \institution{University of Applied Sciences Hamm-Lippstadt}
  \city{Hamm}
  \country{Germany}
}
\email{michellecelina@web.de	}

\author{Maximilian Warsinke}
\orcid{0009-0004-0264-5619}
\affiliation{%
 \institution{Quality and Usability Lab \\Technische Universität Berlin}
 \city{Berlin}
 \country{Germany}}
\email{warsinke@tu-berlin.de}

\author{Adriana-Simona Mihaita}
\orcid{0000-0001-7670-5777}
\affiliation{%
  \institution{University of Technology in Sydney}
  \city{Sydney}
  \country{Australia}}
\email{Adriana-Simona.Mihaita@uts.edu.au}

\author{Jan-Niklas Voigt-Antons}
\orcid{0000-0002-2786-9262}
\affiliation{%
  \institution{University of Applied Sciences Hamm-Lippstadt}
  \city{Hamm}
  \state{Nordrhein-Westfalen}
  \country{Germany}
}
\email{Jan-Niklas.Voigt-Antons@hshl.de}

\renewcommand{\shortauthors}{Vona et al.}
\renewcommand{\shorttitle}{Unmanned Aerial Vehicles Control in a Digital Twin}

\begin{abstract}
Controlling Unmanned Aerial Vehicles (UAVs) is a cognitively demanding task, with accidents often arising from insufficient situational awareness, inadequate training, and bad user experiences. Providing more intuitive and immersive visual feedback—particularly through Digital Twin technologies—offers new opportunities to enhance pilot awareness and the  overall experience quality. \color{black}In this study, we investigate how different virtual points of view (POVs) influence user experience and performance during UAV piloting in Virtual Reality (VR), utilizing a digital twin that faithfully replicates the real-world flight environment. We developed a VR application that enables participants to control a physical DJI Mini 4 Pro drone while immersed in  a digital twin with \color{black} four distinct camera perspectives: Baseline View (static external), First Person View, Chase View, and Third Person View. Nineteen participants completed a series of ring-based obstacle courses from each perspective. In addition to objective flight data, we collected standardized subjective assessments of user experience, presence, workload, cybersickness, and situational awareness.  Quantitative analyses revealed that the First Person View was associated with significantly higher mental demand and effort, greater trajectory deviation, but smoother control inputs compared to the Third Person and Chase perspectives. Complementing these findings, preference data indicated that the Third Person View was most consistently favored, whereas the First Person View elicited polarized reactions. \color{black}
\end{abstract}

\ccsdesc[300]{Human-centered computing~Virtual reality}
\ccsdesc[300]{Human-centered computing~User studies}

\keywords{Unmanned Aerial Vehicles, Digital Twin, Virtual Reality, User Experience}
\begin{teaserfigure}
\centering
  \includegraphics[width=\textwidth,height=0.3\textheight]{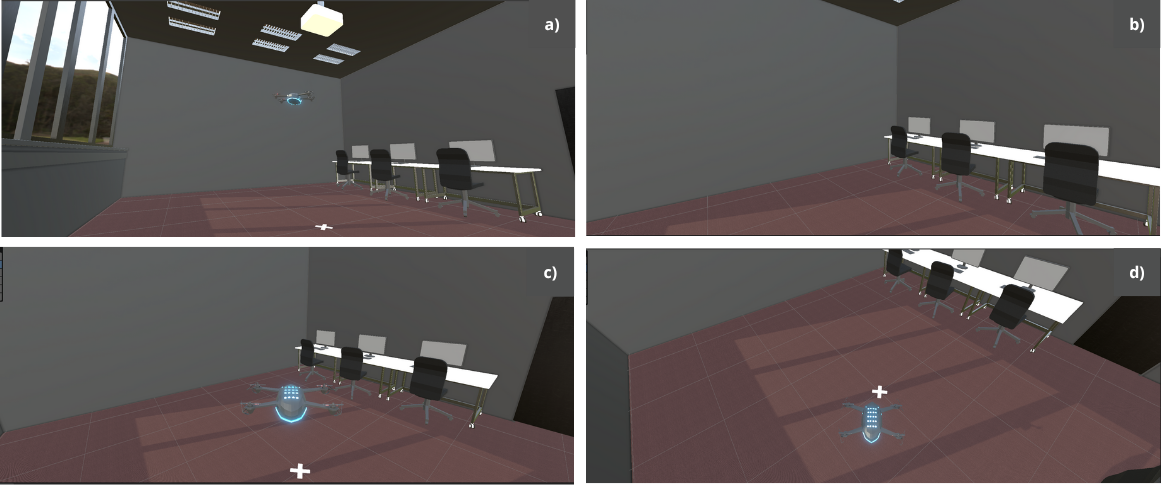}
\caption{Virtual camera perspectives for UAV control in VR: a) Baseline (external pilot viewpoint); b) First Person (front of drone); c) Third Person (elevated behind, wide view); d) Chase (directly behind, following drone).}
  \Description{Virtual camera perspectives used as experimental conditions for UAV control in Virtual Reality: a) Baseline View (static external perspective, approximating the physical pilot viewpoint); b) First Person View (FPV, camera positioned at the front of the drone); c) Third Person View (elevated, slightly behind and above the drone, providing a wide perspective on both the drone and the surrounding environment); d) Chase View (camera positioned directly behind the drone, following its movement).}
  \label{fig:teaser}
\end{teaserfigure}


\maketitle

\section{Introduction}
Unmanned Aerial Vehicles (UAVs), commonly known as drones, have rapidly become indispensable across diverse domains such as aerial photography, infrastructure inspection, agriculture, and search-and-rescue operations~\cite{weibel2005safety}. With expanding application areas, optimizing the overall user experience (UX) of UAV control systems has become a central challenge, particularly as novice pilots and non-experts increasingly engage with these technologies. Although automation and Artificial Intelligence (AI) are advancing, real-world UAV piloting—especially in cluttered or near-earth environments—often demands intuitive interfaces that support rapid human decision-making and reduce the risk of error~\cite{weibel2005safety}.
Traditional ground-based UAV control stations typically offer pilots only limited perspectives—such as a fixed external view or a forward-facing camera—potentially restricting spatial awareness, immersion, and usability. This limitation can impair not only performance and safety but also the pilot’s subjective experience of control and engagement ~\cite{larochelle2012multi,drury2006comparing}. In response, the integration of Virtual Reality (VR) and Digital Twin technologies is emerging as a promising direction for enriching the operator’s experience. Digital Twins—virtual replicas of physical environments—can synchronize in real-time with UAV flight, allowing immersive visualization and interaction within a high-fidelity simulation of the real world ~\cite{lan2016development}.
A crucial element of effective UAV control is situational awareness (SA), which encompasses the pilot’s perception, comprehension, and projection of relevant elements in the environment~\cite{endsley1988design}. Although SA is often discussed in terms of safety and mission success, it is also a key contributor to overall user experience: pilots who feel more aware and in control are likely to rate their experience as more positive and effective~\cite{endsley2011design,helle2022situational,oury2021how}. Within immersive digital twin environments, the way visual information is presented—specifically, the choice of camera perspective—directly shapes both the user’s ability to monitor the drone and its surroundings (supporting SA) and their subjective sense of presence, comfort, and control (contributing to UX). Different perspectives, such as first-person, chase, or third-person views, offer varying amounts of environmental context, self-location cues, and immersion, which can facilitate or hinder both situational awareness and user experience. Despite their importance, the relative benefits and trade-offs of these camera perspectives for UAV piloting in virtual environments have not been systematically examined. This motivates a targeted investigation into how alternative viewpoints in a digital twin setting influence both SA and UX, aiming to identify interface configurations that best support intuitive, effective drone control. To address this gap, we present a controlled laboratory study systematically comparing four distinct virtual camera perspectives—Baseline (static external view), First Person View (FPV), Chase View (behind the UAV), and Third Person View (elevated behind the UAV)—in the context of UAV piloting through a synchronized digital twin. Our contribution is threefold: (1) we introduce a custom VR interface tightly coupled with a real DJI Mini 4 Pro drone in a high-fidelity digital twin, enabling simultaneous physical–virtual control; (2) we advance UAV evaluation by combining objective flight performance metrics, including Dynamic Time Warping (DTW) trajectory analysis, with standardized multidimensional measures of user experience, presence, workload, cybersickness, and situational awareness; and (3) we provide empirical insights and design recommendations on how perspective choice shapes user experience and performance, offering actionable guidance for next-generation UAV control systems that leverage VR and digital twin technologies.
\color{black}

\section{Related Work}
Research on human-drone interaction (HDI) has advanced significantly over the last decade, encompassing topics such as operator workload, collaboration, cognitive support, and the design of intuitive user interfaces~\cite{tezza2019state,ljungblad2021practice}. Surveys of professional drone pilots reveal that practical expertise is shaped not only by technical skill but by experiential factors including situational context, mission type, and cognitive demands~\cite{ljungblad2021practice,peksa2024review,ayala2024uav}. Recent studies have stressed the importance of user-centered design for drone control, identifying usability, learnability, and perceived safety as key determinants of operator satisfaction and system acceptance~\cite{hing2009development,hing2010indoor,zhi2020design}. The integration of intuitive control alternatives, such as gesture-based or head-synced input, is being explored to improve both the accessibility and engagement of drone interfaces ~\cite{watanabe2020head,ribeiro2021uavforeveryone,mahayuddin2017comparison}. Operator experience is further shaped by the type of task and the complexity of the environment. For instance, Larochelle et al.~\cite{larochelle2012multi} and Hing et al.~\cite{hing2009development,hing2010indoor} highlight the demands placed on users during urban search-and-rescue (USAR) or near-earth flight, where rapid decision-making and robust spatial awareness are essential. Studies on remote sensing, inspection, and even spray painting via autonomous UAVs~\cite{mixedrealityfiresmoke2022,spraypainting2017} reinforce the growing breadth of drone applications and the need for interfaces that adapt to diverse user populations.
The camera perspective or point of view (POV) presented to UAV operators has a pronounced effect on spatial orientation, navigation performance, and situational awareness~\cite{hing2010indoor,drury2006comparing,hing2009development}. Multiple studies compare first-person (FPV), chase, and third-person camera perspectives, showing that each offers distinct trade-offs for immersion, awareness, and performance. Hing et al.~\cite{hing2010indoor} find that chase or elevated third-person views can enhance an operator’s ability to perceive spatial relationships and obstacles, thus supporting safer flight in cluttered environments. By contrast, FPV often increases immersion and piloting engagement, but may lead to greater spatial disorientation or collision risk ~\cite{tezza2022first,zhi2020design,mahayuddin2017comparison}. 
Further, the choice of camera perspective can influence pilot workload and learning curves. Drury et al.~\cite{drury2006comparing} demonstrate that multi-view or adaptive perspectives help operators maintain higher situational awareness and task performance, especially during complex missions. These findings are echoed by simulation and survey research investigating user preferences and error rates in different POV configurations ~\cite{larochelle2012multi,mixedrealityfiresmoke2022,pavlik2020drones,hing2009improving}.
Virtual reality (VR), digital twins, and mixed or augmented reality (MR/AR) technologies are increasingly being adopted to enhance UAV piloting and training~\cite{zhi2020design,lan2016development,selecky2017mixed,ribeiro2021web,nielsen2007ecological,virtualenvaerialmonoslam2021,seleky2019analysis}. Digital twins enable precise synchronization of physical UAVs with virtual environments, allowing pilots to practice and operate with a high level of context and realism~\cite{hing2010indoor,selecky2017mixed}. VR-based control platforms can significantly improve usability, perceived presence, and workload reduction by immersing the operator in a first- or third-person digital replica of the flight space~\cite{zhi2020design,lan2016development,selecky2017mixed}.
MR and AR interfaces have also been shown to benefit search and rescue missions, infrastructure inspection, and collaborative UAV operations, supporting incremental development and remote supervision~\cite{humansearchrescue2019,mixedrealityfiresmoke2022,selecky2017mixed,seleky2019analysis}. Usability and safety can be further improved by integrating real-time feedback and multi-sensory cues, as demonstrated in both professional pilot settings and experimental evaluations~\cite{ribeiro2021web,larochelle2012multi}. Notably, mixed reality can also be applied to UAV path planning and autonomous mission design, further expanding the capabilities of modern drone interfaces~\cite{vrpathplanning2021}.
Simulated environments and digital twins provide a safe and controlled means for UAV pilot training, supporting the acquisition of both manual flight skills and situational judgment~\cite{uavsimtraining2019,laTorre2016workload,zhi2020design}. Training research has examined the effects of environmental complexity, instructional design, and scenario variability on workload perception and learning outcomes~\cite{laTorre2016workload,uavsimtraining2019}. Modern simulators facilitate not only skill development but also systematic usability evaluation and comparative testing of interface alternatives.
Recent studies further highlight the impact of information presentation, control modalities, and feedback strategies on pilot workload, learning curves, and user satisfaction~\cite{zhi2020design,nielsen2007ecological,hing2009improving}. Simulator Sickness and cybersickness remain important considerations in VR-based UAV operation, motivating ongoing research into interface adjustments (such as head-synced controls or information overlays) to reduce discomfort and improve accessibility~\cite{watanabe2020head,spraypainting2017,virtualenvaerialmonoslam2021}.
In summary, a large body of work emphasizes the need for user-centered, context-sensitive, and adaptive drone interfaces. Prior studies show that camera perspective, immersive technology, and interface design jointly shape user experience, situational awareness, and training outcomes. However, few studies have directly compared a comprehensive set of virtual camera perspectives within VR-based digital twin systems for real UAV control—particularly in laboratory settings that combine objective flight metrics and subjective UX ratings. The present study addresses this gap with a systematic, multi-perspective evaluation of user experience and performance in immersive UAV piloting.
\section{Methods}
\subsection{Study Design}
The study adopted a within-subjects design to evaluate the effects of four distinct virtual camera perspectives on user experience, situational awareness, and performance during real-world UAV piloting in virtual reality. Each participant completed four different flight parcours under every camera condition: Baseline View (static external), First Person View (FPV), Chase View (directly behind the drone), and Third Person View (elevated, slightly behind and above the drone) (Figure~\ref{fig:teaser}). The sequence of perspectives and parcours was randomized across participants to control for order effects.
The study protocol received approval from the Institutional Ethics Committee. 
\begin{figure}[ht]
    \centering
    \includegraphics[width=0.8\linewidth]{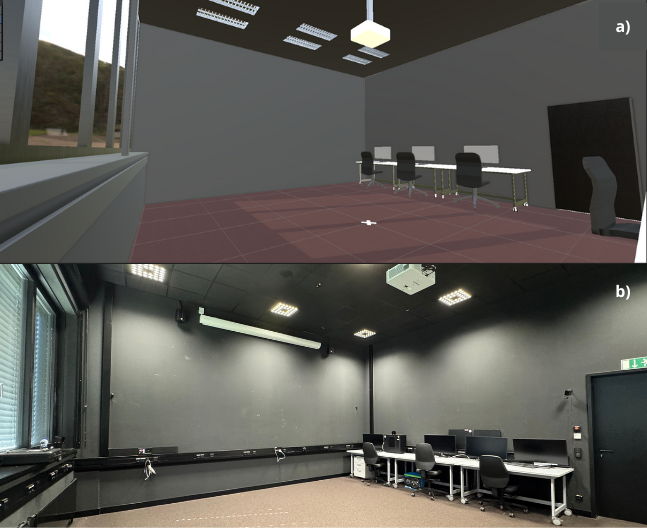}
    \caption{
    Comparison of the laboratory flight environment and its digital twin. 
    \textbf{a)} Virtual model of the laboratory as rendered in the VR environment (digital twin). 
    \textbf{b)} Photograph of the corresponding real laboratory used for UAV experiments.
    }
    \Description{This figure shows a side-by-side comparison between the digital twin (a) and the real-world laboratory (b) where the study was conducted. The digital twin environment (a) was designed to closely match the physical dimensions and layout of the real room (b), supporting high-fidelity, immersive piloting in the virtual reality experiment.}
    \label{fig:digitalrealroom}
\end{figure}

\subsection{Apparatus and Experimental Setup}
The experimental setup was designed around a high-fidelity digital twin of the laboratory environment, enabling synchronized operation between a real drone and its virtual counterpart (Figure~\ref{fig:digitalrealroom}). 

\textbf{Drone and Tracking System.} A DJI Mini 4 Pro drone was used for all experimental flights. The tracking system (4 ARTTRACK5 trackers) operated at 120~Hz with sub-millimeter positional accuracy (±0.2~mm) and a latency of approximately 8~ms. Jitter was below 1~mm during static trials.  However, end-to-end latencies such as pose-to-photon delay in the HMD rendering pipeline and controller-to-drone transmission times were not systematically measured during the study.   The DJI Mini 4 Pro drone was equipped with reflective markers (12~mm) for tracking, and collision events with obstacles were logged both in the physical room and within the Unity simulation. The ART system was operated with the \textit{DTrack} software suite, which provided high-precision position and orientation data.  In the experiment, the drone was flown in normal flight mode across all conditions, with Advanced Pilot Assistance Systems (APAS) enabled and the Brake mode used to stop the drone when approaching obstacles. The physical environment was well-lit, and during testing, the DJI Mini 4 Pro provided reliable and stable object detection.  

\textbf{Head-Mounted Display.} Participants experienced the virtual environment through an HTC Vive Pro 2 head-mounted display (HMD), supported by four HTC base stations for spatial tracking. This setup allowed users to perceive the virtual twin while simultaneously piloting the drone. The HTC Vive Pro 2 headset featured, a refresh rate of 120~Hz, and a field of view of 120$^{\circ}$. 

\textbf{Unity VR Application.} The digital twin was implemented as a desktop virtual reality application developed in Unity~2022.3 LTS. A dedicated plugin enabled direct communication between Unity and the ART software, allowing continuous synchronization of drone position and orientation between the real and virtual environments.
 
The digital twin also comprised the drone model and four predefined flight parcours, each consisting of nine suspended rings that participants had to navigate in sequence. The parcours were designed to vary direction changes and difficulty while maintaining comparable length and number of gates. For example, Parcours~1 required the sequence “up, forward, up, left, left, left, down, right, down,” while Parcours~2 involved “up, right, up, right, right, down, forward, right, down.” Parcours~3 followed “up, left, right, up, right, down, forward, right, down,” and Parcours~4 consisted of “up, forward, right, up, right, down, right, left, down.” At each step, only the next required waypoint was visible to the participant, while subsequent rings remained hidden until reached. This ensured that participants followed the intended sequence of maneuvers and maintained focus on immediate navigation rather than planning the entire trajectory in advance.

\textbf{Camera Perspectives.}  
In all four views, the camera followed the rotation of the drone, except for the Baseline (default) view. In this condition, the camera’s rotation was independent of the drone, and its position was fixed in the corner of the physical and virtual room at 1.1\,m height, 1.9\,m anterior, and 1.9\,m lateral to the drone (Figure\ref{fig:teaser}a). This setup simulated the external viewpoint of a real-world pilot, allowing participants to freely rotate their heads via the HMD. For the other three views, participants were instructed not to rotate their bodies but instead to rotate the drone, while still being free to move their heads. In all conditions, the field of view was set to 60$^{\circ}$.  
\textit{First-Person View.} The camera occupied the same position and orientation as the drone, following its rotation and translation. The drone body itself was not visible in this view (Figure\ref{fig:teaser}b).  
\textit{Chase View.} The camera was positioned 1.695\,m above and 0.86\,m behind the drone, tilted downward at 60$^{\circ}$. When the participant looked forward, the resulting view emphasized the area in front of the drone while preserving strong situational awareness of the environment (Figure\ref{fig:teaser}d).  
\textit{Third-Person View.} The camera was placed 0.9\,m behind and 0.4\,m above the drone, tilted downward at 12$^{\circ}$. In this condition, participants could see both the drone and parts of its immediate surroundings (Figure\ref{fig:teaser}c).  
To ensure a natural viewing experience, camera position and rotation were smoothed using linear interpolation (\texttt{Vector3.Lerp}) and spherical linear interpolation (\texttt{Quaternion.Slerp}), respectively. Interpolation factors were scaled by \texttt{Time.deltaTime} and multiplied by smoothing speeds of 30 (position) and 20 (rotation). This minimized abrupt transitions, providing stable but responsive camera movements without noticeable lag.

\textbf{Calibration Procedure.} At the beginning of each session, the drone was placed at the center of the physical room to perform calibration within the ART tracking system. Once calibration confirmed correct alignment, the Unity application was launched on the desktop PC, and the HTC Vive Pro 2 headset was initialized. This ensured that the participant could pilot the drone both in the real flight room and in the virtual twin environment contemporaneously, with matched dynamics and obstacle layout.
\color{black}
\subsection{Procedure}
Participants were first welcomed into the laboratory, provided with a study overview, and completed an informed consent process. Baseline demographic data and background information (e.g., age, gender, prior XR and drone experience, technology affinity) were collected via an initial questionnaire.
Participants were then introduced to the VR system, which comprised an HTC Vive Pro headset and a physical DJI remote controller. A standardized tutorial session guided each participant through the drone control mechanics and VR interface navigation. This tutorial included a brief free-flight period in the digital twin environment, allowing participants to practice basic maneuvers and become familiar with the virtual setting and control scheme.
Following the tutorial, participants completed four flight tasks, one per camera perspective, in a randomized order. For each task, the participant donned the VR headset and navigated the drone through a series of suspended rings, attempting to fly as centrally and efficiently as possible. The digital twin ensured that all visual cues and environmental obstacles in VR matched their real-world counterparts. Participants interacted directly with the physical remote controller throughout, while the VR system displayed the live-synced digital twin from the current perspective.
After each flight session, participants removed the headset and completed a set of questionnaires evaluating user experience, presence, workload, cybersickness, and situational awareness for that perspective. Upon finishing all four conditions, participants ranked their preferred and least preferred perspectives and provided qualitative comments on their experience.
 Safety measures included monitoring for simulator sickness by regularly checking participant well-being, providing clear exit instructions, and allowing participants to withdraw from the experiment at any time without penalty. Equipment was sanitized before each session to maintain hygiene standards. An experimenter was present at all times to intervene if participants experienced discomfort or disorientation.
\color{black}
\subsection{Questionnaires, Variables, and Data Analysis}
The study employed a combination of subjective questionnaires to capture different facets of user experience and performance. The \textbf{User Experience Questionnaire Short (UEQ-S)} provided pragmatic and hedonic quality scores as well as an overall total score (UEQ\_PRAG, UEQ\_HED, and UEQ\_TOT) \cite{ueqs,ueqscenario}. The \textbf{Igroup Presence Questionnaire (IPQ)} yielded scores for total presence, General presence (IPQ\_G), Spatial presence (IPQ\_SP), Involvement (IPQ\_INV), and Realism (IPQ\_REAL)~\cite{schubert2003sense,tran2024survey,regenbrecht2022designing}. Perceived workload was measured using the \textbf{NASA Task Load Index (NASA-TLX)}, which provided a total workload score (NASA\_TOT)~\cite{hart2006nasa}. Symptoms of simulator sickness were assessed with the \textbf{Cybersickness Questionnaire (CSQ-VR)}, which provided subscale scores for nausea (CSQ\_NAU), vestibular symptoms (CSQ\_VES), and oculomotor symptoms (CSQ\_OCU), as well as an overall total score (CSQ\_TOT)~\cite{virtualworlds2010002}. Situation awareness was measured using the \textbf{Situation Awareness Rating Technique (SART)}, with a total score (SART\_TOT) and three subscales: Demand (SART\_D), Supply (SART\_S), and Understanding (SART\_U). The validated 9-item version was used, omitting the “Unpredictability of the situation” item in line with prior recommendations~\cite{taylor1990sart,selcon1991,mica1998,wayne1994tools,nguyen2019,zhou2022}. Finally, technology acceptance was assessed with the \textbf{Affinity for Technology Interaction (ATI) scale}, which measured participants’ openness and acceptance of new technologies as a mean ATI score~\cite{franke2019affinity}.
Objective performance variables included the time to complete each parcours, the position of the drone, and detailed indicators of flight path accuracy derived from Dynamic Time Warping (DTW) analysis. Specifically, we computed DTW Distance, DTW Normalized Distance (accounting for path length), Average and Maximum Deviation from the reference trajectory, and the difference between actual and reference path lengths. All performance data were automatically recorded and linked to the corresponding camera perspective. 
For all performance measures, we conducted both one-way repeated measures ANOVAs and non-parametric Friedman tests to ensure robustness. When omnibus tests indicated significant effects, post-hoc pairwise Wilcoxon signed-rank tests were applied, with Bonferroni or Holm corrections to control for multiple comparisons. Descriptive statistics were computed for all measures, and qualitative responses regarding preferred and least-preferred perspectives were thematically coded to contextualize the quantitative results.

\color{black}
\subsection{Participants}
A total of 21 individuals were recruited for the study; two participants were excluded due to incomplete or missing data, resulting in a final sample of 19 participants (16 male, 3 female; M\textsubscript{age} = 25.0 years, SD = 5.9). The majority reported prior experience with extended reality (XR) technologies (n = 18), though frequency of use varied widely: six participants indicated use less than once per year, three used XR once or twice per year, two once or twice per month, three once or twice per week, and three reported almost daily use. Most participants had spent less than one hour using XR (n = 6), with additional responses spread across higher usage brackets.
Private use was the most commonly reported context for immersive media experience (n = 10), while virtual reality (VR) was the most frequently cited type of immersive technology (n = 10), followed by combinations of VR, AR, and MR (n = 5). On a 7-point scale, the mean self-reported confidence using immersive media was 5.35 (SD = 1.18), and mean competence was 4.70 (SD = 1.38).
Fifteen participants indicated prior experience with videogames, with a variety of game genres represented. Experience with drone flight was evenly distributed (n = 9 yes, n = 9 no), with the most common amount of previous drone exposure being less than one hour (n = 11). Most reported no sensory limitations relevant to the study (n = 16), with two noting limited vision.
Technology acceptance was assessed using the ATI scale, yielding a mean of 4.51 (SD = 0.52) and good internal consistency ($\alpha = .79$).

\section{Results}

\subsection{Workload (NASA-TLX)}
\begin{figure}[!ht]
\centering
\includegraphics[width=\linewidth]{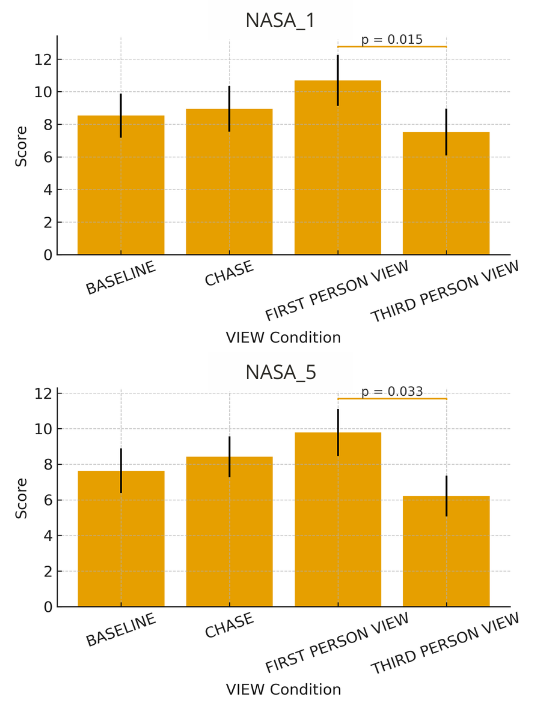}
\caption{NASA-TLX subscales showing significant differences across VIEW conditions. 
Mental Demand (NASA\_1) and Effort (NASA\_5) both indicated higher workload in the FIRST PERSON VIEW compared to the THIRD PERSON VIEW. Error bars represent the standard error of the mean (SEM).}
\label{fig:nasa_subscales}
\end{figure}
Descriptive statistics indicated that perceived workload, as measured by the NASA-TLX, was lowest for the THIRD PERSON VIEW condition (M = 38.00, SD = 24.79), followed by BASELINE (M = 43.53, SD = 24.43) and CHASE (M = 45.11, SD = 24.70), and highest for the FIRST PERSON VIEW condition (M = 52.89, SD = 31.93).
 Mauchly’s test indicated that the assumption of sphericity was violated for NASA-TLX total scores ($W = 0.069$, $p < .001$). Therefore, Greenhouse–Geisser corrections were applied ($\varepsilon = .57$). The corrected ANOVA did not reveal a significant effect of VIEW, $F(1.72, 30.99) = 2.81$, $p = .083$. 
  In contrast, a non-parametric Friedman test confirmed a significant main effect, $\chi^2(3) = 10.18$, $p = .017$, suggesting differences in workload across perspectives.
To explore which conditions drove this effect, post-hoc Wilcoxon signed-rank tests with Holm correction were conducted. However, none of the pairwise comparisons reached statistical significance after correction, suggesting that with the present sample size, the overall effect cannot be attributed to specific condition differences with sufficient confidence. 
For the six NASA-TLX subscales (NASA\_1–6), Mauchly’s tests likewise indicated violations of sphericity (all $p < .001$). Greenhouse–Geisser corrected ANOVAs revealed significant effects of VIEW for NASA\_2, $F(1.58, 28.41) = 3.90$, $p = .041$, and NASA\_5, $F(2.04, 36.65) = 3.37$, $p = .045$, as well as a trend-level effect for NASA\_1, $F(1.59, 28.68) = 2.72$, $p = .094$. No significant effects were observed for NASA\_3, NASA\_4, or NASA\_6 (all $p > .17$). These findings were supported by non-parametric Friedman tests, which confirmed significant VIEW effects for NASA\_1, $\chi^2(3) = 9.13$, $p = .028$, and NASA\_2, $\chi^2(3) = 11.09$, $p = .011$, and indicated a trend for NASA\_5, $\chi^2(3) = 7.02$, $p = .071$.
Post-hoc Wilcoxon signed-rank tests with Holm correction indicated that workload was significantly higher in the FIRST PERSON VIEW compared to the THIRD PERSON VIEW for NASA\_1 ($p = .015$) and NASA\_5 ($p = .033$). No other pairwise comparisons reached significance after correction.
Figure~\ref{fig:nasa_subscales} illustrates the two NASA-TLX subscales that showed significant differences across VIEW conditions, with both Mental Demand and Effort indicating higher workload in the FIRST PERSON VIEW compared to the THIRD PERSON VIEW.
\color{black}
\subsection{User Experience (UEQ)}

Descriptive statistics for the User Experience Questionnaire subscales showed that, for the pragmatic quality subscale (UEQ\_PRAG), mean scores were highest for THIRD PERSON VIEW (M = 1.95, SD = 0.93), followed by BASELINE (M = 1.53, SD = 0.92), CHASE VIEW (M = 1.49, SD = 1.34), and lowest for FIRST PERSON VIEW (M = 1.03, SD = 1.89). For the hedonic quality subscale (UEQ\_HED), the highest mean score was also observed for THIRD PERSON VIEW (M = 1.82, SD = 0.94). The overall UEQ score (UEQ\_TOT) was likewise highest for THIRD PERSON VIEW (M = 1.88, SD = 0.81), with the lowest value again in FIRST PERSON VIEW (M = 1.28, SD = 1.44).
 Mauchly’s tests indicated that the assumption of sphericity was violated for all UEQ measures (all $p < .001$). Therefore, Greenhouse–Geisser corrections were applied. The corrected ANOVA for UEQ\_TOT did not reveal a significant effect of VIEW, $F(1.62, 29.23) = 2.51$, $p = .108$, consistent with the non-parametric Friedman test, $\chi^2(3) = 4.55$, $p = .208$. For the pragmatic quality subscale, the Greenhouse–Geisser corrected ANOVA showed a trend-level effect of VIEW, $F(1.81, 32.63) = 3.00$, $p = .068$, while the Friedman test indicated a significant effect, $\chi^2(3) = 8.83$, $p = .032$. No effect of VIEW was found for the hedonic quality subscale in either analysis ($p > .36$).
 Post-hoc Wilcoxon signed-rank tests with  Holm correction did not identify any statistically significant pairwise differences between VIEW conditions for UEQ\_PRAG, indicating that while pragmatic quality varied across conditions overall, no specific pairwise comparison survived correction.
\color{black}
\subsection{Cybersickness (CSQ-VR)}
Descriptive analysis of the Cybersickness Questionnaire revealed clear differences in self-reported nausea (CSQ\_NAU) across drone piloting perspectives. Mean nausea scores were highest in the FIRST PERSON VIEW condition (M = 5.32, SD = 4.38), followed by THIRD PERSON VIEW (M = 4.16, SD = 3.11), CHASE VIEW (M = 4.11, SD = 3.13), and lowest in the BASELINE condition (M = 3.53, SD = 2.65). For the Vestibular subscale (CSQ\_VES), mean scores were M = 3.95 (SD = 3.14) for BASELINE, M = 4.84 (SD = 3.32) for CHASE VIEW, M = 5.79 (SD = 3.91) for FIRST PERSON VIEW, and M = 4.37 (SD = 3.22) for THIRD PERSON VIEW. On the Oculomotor subscale (CSQ\_OCU), mean scores were M = 3.74 (SD = 2.60) for BASELINE, M = 3.95 (SD = 2.82) for CHASE VIEW, M = 5.16 (SD = 3.99) for FIRST PERSON VIEW, and M = 4.37 (SD = 3.40) for THIRD PERSON VIEW. The overall CSQ total score (CSQ\_TOT) was also highest in FIRST PERSON VIEW (M = 16.26, SD = 11.85), followed by CHASE VIEW (M = 12.89, SD = 8.28), THIRD PERSON VIEW (M = 12.89, SD = 9.39), and BASELINE (M = 11.21, SD = 7.42).
Mauchly’s test indicated violations of sphericity for all CSQ measures ($p < .001$). Accordingly, Greenhouse–Geisser corrected ANOVAs were applied. For nausea, the ANOVA effect of VIEW was no longer significant after correction, $F$(1.51, 27.25) = 2.81, $p = .090$, whereas the non-parametric Friedman test confirmed a significant effect, $\chi^2$(3) = 8.94, $p = .030$. No significant effects of VIEW were observed for the vestibular, oculomotor, or total CSQ scores in either ANOVA (all $p > .10$) or Friedman tests (all $p > .09$).  Post-hoc Wilcoxon signed-rank tests with Holm correction did not reveal any statistically significant pairwise differences, although the lowest corrected $p$-value was observed between BASELINE and FIRST PERSON VIEW ($p = .068$).

\color{black}
\subsection{Presence (IPQ)}
Descriptive statistics for the Igroup Presence Questionnaire (IPQ) indicated that mean total presence scores (IPQ\_TOT) were similar across all VIEW conditions, with values ranging from M = 3.53, SD = 0.62 for CHASE VIEW to M = 3.73, SD = 0.57 for THIRD PERSON VIEW. General Presence (IPQ\_G) also showed comparable means across conditions (M = 4.58, SD = 1.02 for BASELINE to M = 4.26, SD = 1.41 for THIRD PERSON VIEW), as did Involvement (IPQ\_INV: M = 3.39, SD = 0.96 for CHASE VIEW to M = 3.71, SD = 1.01 for THIRD PERSON VIEW) and Realism (IPQ\_REAL: M = 2.83, SD = 0.86 for FIRST PERSON VIEW to M = 3.25, SD = 0.46 for BASELINE). The Spatial Presence subscale (IPQ\_SP) was highest for FIRST PERSON VIEW (M = 4.34, SD = 1.04), followed by THIRD PERSON VIEW (M = 4.16, SD = 0.64), BASELINE (M = 4.08, SD = 0.79), and CHASE VIEW (M = 3.88, SD = 0.80).
Mauchly’s tests indicated violations of the sphericity assumption for all subscales ($p < .001$); therefore, Greenhouse–Geisser corrections were applied. Repeated measures ANOVAs revealed no significant effects of VIEW on total presence ($F$(1.60, 28.82) = 0.89, $p = .40$), general presence ($F$(1.62, 29.23) = 0.44, $p = .64$), involvement ($F$(1.60, 28.82) = 0.67, $p = .51$), or realism ($F$(1.59, 28.41) = 1.75, $p = .19$). For spatial presence (IPQ\_SP), the GG-corrected ANOVA was not significant ($F$(1.60, 28.82) = 0.89, $p = .40$), but the non-parametric Friedman test indicated a significant main effect of VIEW, $\chi^2$(3) = 9.32, $p = .025$. 
However, post-hoc Wilcoxon signed-rank tests with Holm correction did not reveal any significant pairwise differences, suggesting that although spatial presence varied across conditions overall, no specific VIEW comparison could be localized with sufficient statistical confidence.

\color{black}
\subsection{Situational Awareness (SART)}
Descriptive statistics for the SART revealed similar situational awareness scores across VIEW conditions, with mean SART\_TOT ranging from M = 9.58, SD = 5.07 for CHASE to M = 11.47, SD = 5.19 for FIRST PERSON VIEW. Subscale analyses showed that Understanding (SART\_U) and Supply (SART\_S) were generally highest in FIRST PERSON VIEW (M\_{U} = 7.11, SD = 3.20; M\_{S} = 12.74, SD = 6.56), while Demand (SART\_D) was also highest in FIRST PERSON VIEW (M = 8.37, SD = 6.55) and lowest in BASELINE (M = 5.42, SD = 4.02).
 Mauchly’s tests indicated violations of the sphericity assumption for all subscales ($p < .001$); therefore, Greenhouse–Geisser corrections were applied. The corrected repeated measures ANOVAs revealed no significant main effects of VIEW on total situational awareness ($F$(1.60, 28.72) = 1.82, $p = .19$), Understanding ($F$(1.60, 28.72) = 1.82, $p = .19$), Demand ($F$(1.60, 28.72) = 1.82, $p = .19$), or Supply ($F$(1.60, 28.72) = 1.82, $p = .18$). Consistently, non-parametric Friedman tests did not indicate significant differences between VIEW conditions for any measure (all $p > .05$). 
 Follow-up Wilcoxon tests on the subscales revealed no pairwise differences that survived Holm correction.

\subsection{Participant Preferences}

In addition to standardized questionnaires, participants (N = 19) indicated their most and least preferred camera perspectives and provided short written justifications. A frequency analysis revealed differences across conditions. The THIRD PERSON VIEW was the most frequently preferred perspective (9 participants) and was notably never selected as the least preferred. In contrast, the FIRST PERSON VIEW was polarizing: while 6 participants listed it as their top preference, it was also the most frequently disliked perspective (9 participants). The CHASE VIEW was chosen as top preference by only 3 participants, but as least preferred by 6, while the BASELINE VIEW was rarely preferred (1 participant) and often rated least preferred (4 participants).
Representative comments highlight the reasoning behind these choices. Participants who favored the THIRD PERSON VIEW emphasized clarity and control, noting it offered ``easy control'' (P3) and allowed them to ``see [the] majority of the environment'' (P4). In contrast, those who preferred the FIRST PERSON VIEW valued immersion and control, describing that they ``felt fully immersed in the drone’s perspective'' (P1) or that ``I felt like I am the drone and more in control'' (P2). However, others found the same perspective difficult, reporting that it was ``very confusing'' (P3), ``could not control the drone well'' (P2), or ``physically demanding'' (P4). The CHASE VIEW was criticized as ``least immersive and uncomfortable'' (P1), and the BASELINE VIEW was described as less engaging, with one participant noting ``I never did it before and I feel comfortable [with other views]'' (P5).

\color{black}
\begin{table*}[!ht]
\centering
\caption{Summary of significant post-hoc comparisons across VIEW conditions. Only pairwise effects that survived correction are reported.}
\label{tab:posthoc_results}
\resizebox{\textwidth}{!}{%
\begin{tabular}{llp{10cm}}
\hline
\textbf{Measure} & \textbf{Subscale / Metric} & \textbf{Significant Pairwise Differences} \\
\hline
NASA-TLX & Mental Demand (NASA\_1) & FIRST PERSON VIEW $>$ THIRD PERSON VIEW ($p = .015$) \\
         & Effort (NASA\_5)        & FIRST PERSON VIEW $>$ THIRD PERSON VIEW ($p = .033$) \\
         \hline
Performance & DTW Distance          & FIRST PERSON VIEW $>$ CHASE VIEW ($p = .028$ and W=19.0); FIRST PERSON VIEW $>$ THIRD PERSON VIEW ($p = .006$ and W=12) \\
 & Smoothness            & FIRST PERSON VIEW $>$ CHASE VIEW ($p < .05$); FIRST PERSON VIEW $>$ THIRD PERSON VIEW ($p < .05$) \\
\hline
\end{tabular}%
}
\end{table*}
\subsection{Performance}

\begin{figure}[ht]
    \centering
    \includegraphics[width=0.9\linewidth]{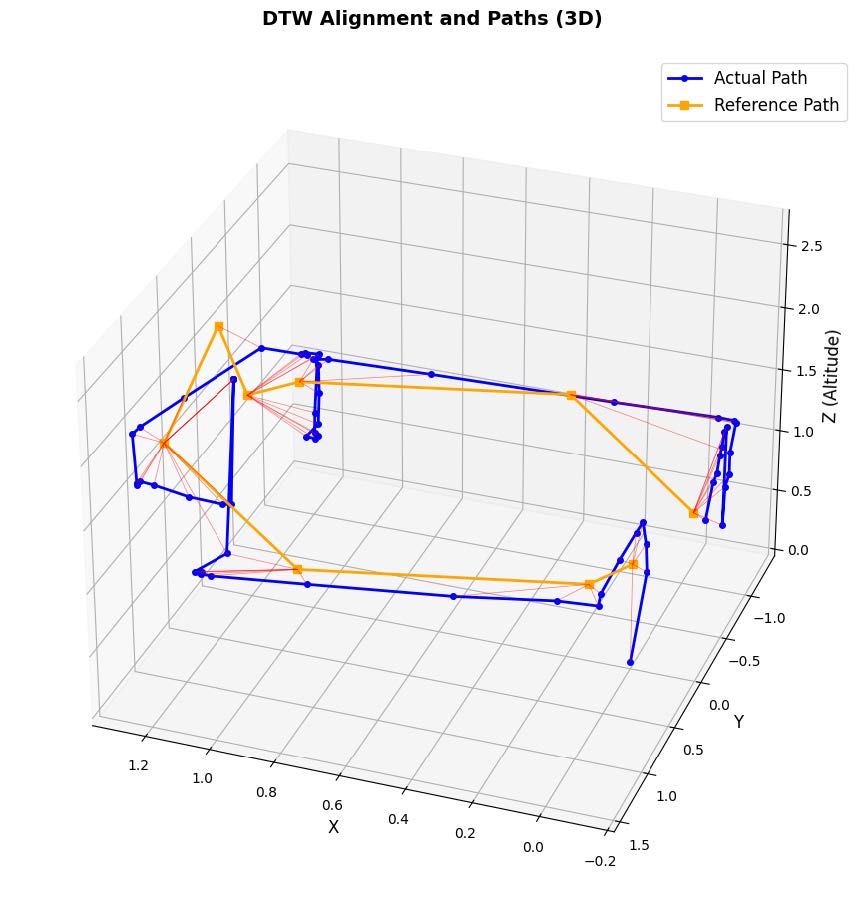}
    \caption{DTW alignment example showing a representative actual drone flight path (blue) and the optimal reference path (orange) in 3D space. Axes represent normalized spatial coordinates.}
    \label{fig:dtw_3d}
\end{figure}

\begin{figure}[ht]
    \centering
    \includegraphics[width=0.9\linewidth]{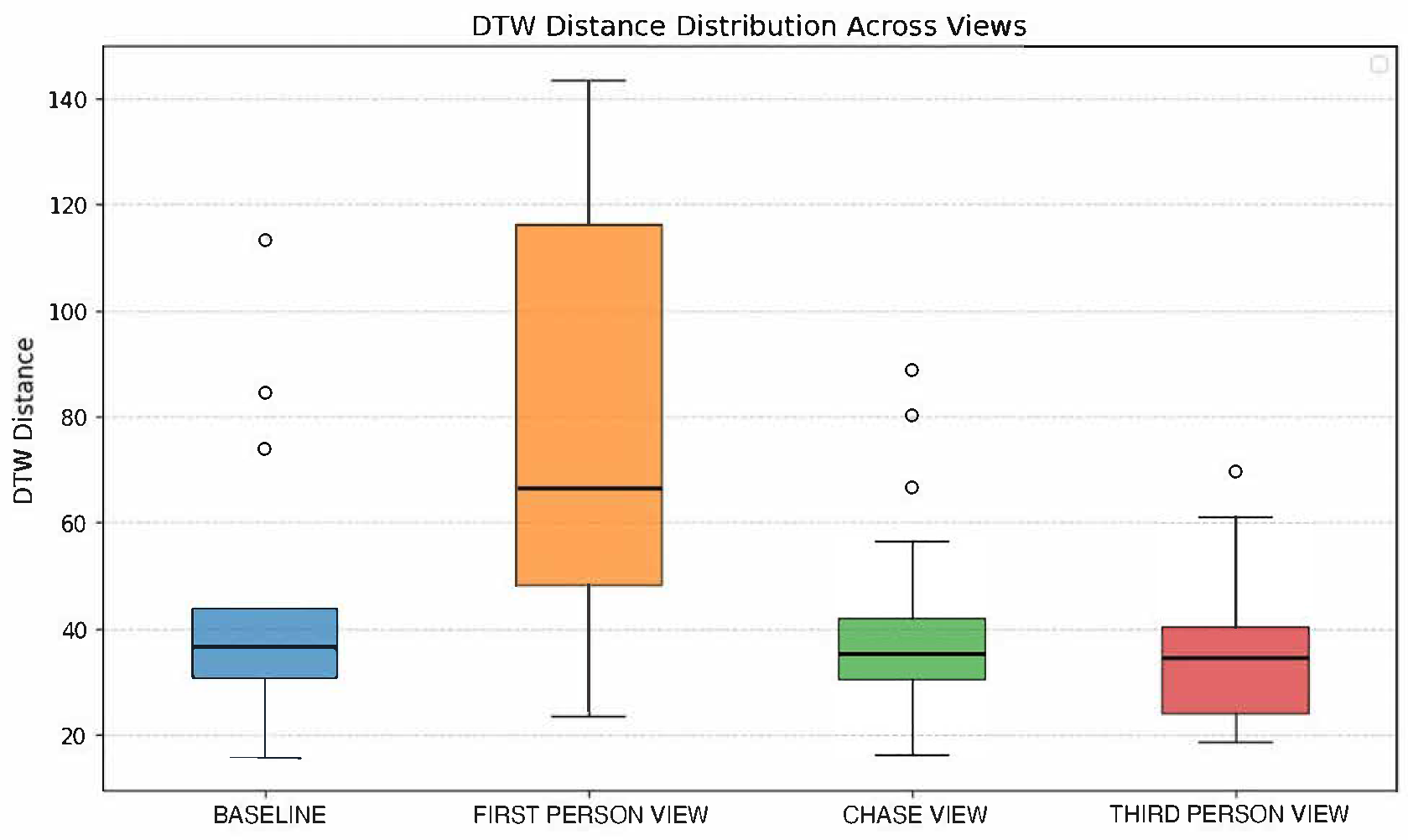}
    \caption{Boxplot of DTW distance distributions across all four camera perspectives.}
    \label{fig:dtw_boxplot}
\end{figure}

To objectively assess piloting performance across camera perspectives, several metrics were extracted from each drone trajectory. These included Dynamic Time Warping (DTW) distance, normalized DTW distance, trajectory smoothness, total flight time, and additional measures of average and maximum deviation from the optimal path. Dynamic Time Warping was implemented using the \texttt{dtw} package~\cite{giorgino2009dtw}, which computes the minimum cumulative distance needed to align the participant’s path with a reference trajectory, thereby providing a robust, stretch-insensitive measure of flight accuracy.
\textbf{Dynamic Time Warping Distance}
The raw DTW distance, representing total deviation from the optimal path (in meters), showed a significant effect of VIEW in both analyses ( Friedman $\chi^{2}(3) = 8.08$, $p = .044$; ANOVA $F(3,48) = 10.77$, $p < .00005$). Mauchly’s test indicated that the assumption of sphericity was violated for this metric ($W = 0.436$, $\chi^{2}(5) = 12.24$, $p = .032$). Therefore, Greenhouse–Geisser corrections were applied ($\varepsilon = 0.63$), which confirmed the significance of the effect ($F(1.90,30.41) = 10.77$, $p = .0003$). \color{black}Post-hoc pairwise Wilcoxon tests with Bonferroni correction indicated that FIRST PERSON VIEW produced significantly higher DTW distances than both CHASE VIEW ($p = .028$, $W = 19.0$) and THIRD PERSON VIEW ($p = .006$, $W = 12$), while no other comparisons reached significance. Median DTW distances were 36.65 for BASELINE, 66.53 for FIRST PERSON VIEW, 35.43 for CHASE VIEW, and 34.59 for THIRD PERSON VIEW, indicating that participants deviated most from the reference path in FIRST PERSON VIEW and least in THIRD PERSON VIEW. The distribution of DTW distances across perspectives is visualized in Figure~\ref{fig:dtw_boxplot}, while Figure~\ref{fig:dtw_3d} illustrates a representative DTW alignment in three-dimensional space.
\textbf{Normalized DTW Distance}
Normalized DTW distance showed a trend-level effect across perspectives ( Friedman $\chi^{2}(3) = 7.94$, $p = .047$; ANOVA $F(3,48) = 2.63$, $p = .061$). Mauchly’s test indicated no violation of sphericity ($W = 0.821$, $p = .715$).\color{black} This suggests that while total deviations differed, the accuracy per unit of path length was broadly comparable across perspectives.
\textbf{Smoothness}
Smoothness, quantified as the mean rate of change of acceleration (jerk), revealed a significant main effect of VIEW in both tests ( Friedman $\chi^{2}(3) = 14.15$, $p = .003$; ANOVA $F(3,48) = 6.98$, $p = .0005$). Mauchly’s test indicated no violation of sphericity ($W = 0.662$, $p = .301$). \color{black}Post-hoc comparisons showed that FIRST PERSON VIEW yielded significantly smoother flights than both CHASE VIEW and THIRD PERSON VIEW (both $p < .05$ after correction). Thus, although FIRST PERSON VIEW was associated with larger deviations from the reference path, it enabled smoother, less abrupt control inputs.
\color{black}
\textbf{Flight Time}
Analysis of total flight duration revealed differences across perspectives ( Friedman $\chi^{2}(3) = 7.73$, $p = .052$; ANOVA $F(3,48) = 10.82$, $p < .00005$). Mauchly’s test indicated that the assumption of sphericity was not violated ($W = 0.519$, $p = .086$). \color{black} Mean (and median) flight times were: BASELINE, 89.59s (76.00s); FIRST PERSON VIEW, 142.88s (126.00s); CHASE VIEW, 86.53s (75.00s); THIRD PERSON VIEW, 76.35s (60.01s). These results indicate that FIRST PERSON VIEW was associated with substantially longer flight durations.
\textbf{Additional Metrics}
Average deviation, maximum deviation, and the difference in path length between the optimal and actual trajectory showed no significant effect of VIEW in either analysis (all $p > .20$). For these metrics, Mauchly’s tests confirmed that the assumption of sphericity was met (all $p > .07$).

\section{Discussion}
 
This study provides a systematic, multi-dimensional comparison of virtual camera perspectives for real-world UAV piloting in a digital twin VR environment. While most effects did not reach statistical significance, we also considered qualitative patterns in the data and compared these with existing literature on immersive drone operation to contextualize our findings~\cite{hing2010indoor,drury2006comparing}.
\color{black}
\subsection{Interpretation in Context of Related Work}

\textbf{Workload and User Experience.} Our clearest evidence emerged from the NASA-TLX subscales, where the FIRST PERSON VIEW was associated with higher Mental Demand and Effort compared to the third-person view. These pairwise effects survived correction and converge with prior work showing that third-person or chase perspectives facilitate more intuitive navigation and reduce task load in UAV contexts~\cite{hing2009development,larochelle2012multi,zhi2020design}.
The FIRST PERSON VIEW, while intuitively appealing and rated highest in immersion and naturalness (as measured by IPQ-SP and CSQ Nausea), was paradoxically associated with higher cognitive workload, more self-reported nausea, and poorer objective flight accuracy (higher DTW distance and longer completion times). This aligns with existing evidence that, while FIRST PERSON VIEW can deepen the sense of presence~\cite{tezza2020lets,pavlik2020drones}, it also places increased demands on spatial memory and control, and can elevate the risk of VR sickness~\cite{watanabe2020head}.
For overall workload and UEQ measures, however, differences between perspectives were less robust and often did not survive conservative correction. This suggests that while viewpoint influences user experience, the direction and strength of effects may vary across individuals and contexts echoing calls for adaptable, user-centered interfaces ~\cite{ljungblad2021practice,tezza2019state}.
\textbf{Cybersickness and Presence.}
Cybersickness effects were not significant across most subscales, but there was a trend toward increased nausea in FIRST PERSON VIEW compared to baseline—a result to consider for training and prolonged use. Sense of presence (IPQ) did not differ significantly across perspectives, indicating that even more “detached” views can deliver immersive experiences if the digital twin is well-designed. This supports prior findings that presence is shaped by more than just camera position~\cite{schubert2003sense,tran2024survey}.
\textbf{Situational Awareness.}
No significant differences were found in overall SART scores or subscales, though there was a trend toward higher understanding and demand in FIRST PERSON VIEW. This finding echoes literature noting that increased immersion can boost environmental engagement but may also amplify task demand~\cite{endsley1988design,selcon1991}.
\textbf{Objective Performance.}
Performance analysis revealed that participants deviated most from the optimal path in FIRST PERSON VIEW, while THIRD PERSON VIEW yielded the best alignment (lowest median DTW distance). THIRD PERSON VIEW and CHASE VIEW also produced more consistent trajectories, while FIRST PERSON VIEW was associated with greater variability and longer completion times. Interestingly, flight smoothness (mean jerk) was better in FIRST PERSON VIEW, possibly indicating slower, more careful movements when participants were less confident or comfortable. This aligns with prior findings that viewpoint can affect not only the accuracy but also the style of UAV control~\cite{hing2009development}.
\textbf{User preferences} Results on user preferences suggest that while the THIRD PERSON VIEW provided a consistently positive experience, the FIRST PERSON VIEW divided participants, offering strong immersion for some but high workload and confusion for others. The CHASE VIEW and BASELINE VIEW were rarely favored and often criticized, indicating limited suitability as primary control perspectives.
\color{black}
\subsection{Design Guidelines and Application Implications}

Our results yield several concrete design recommendations for immersive UAV control systems:

\textbf{Multi-perspective flexibility:} Enable flexible viewpoint switching to accommodate individual preferences and task demands, as prior literature and our results show strong variability in what works best for different users~\cite{ljungblad2021practice}.

\textbf{Default to third-person in cluttered or complex environments:} Third-person views provide an optimal balance of awareness and workload reduction for navigation, training, and general piloting~\cite{hing2009development,hing2010indoor}.

\textbf{FPV as an option:} Offer FPV as an option for scenarios where immersion or naturalistic control is paramount (e.g., simulation, search, or inspection), but supplement it with supportive cues or overlays to mitigate workload and cybersickness risks~\cite{drury2006comparing,nielsen2007ecological}.

\textbf{Monitor for VR sickness and adapt exposure:} Since immersive FPV can increase cybersickness risk, control systems should integrate real-time monitoring (e.g., session timers, adaptive FOV) and provide guidance for breaks~\cite{watanabe2020head}.

\subsection{Limitations and Future Work}
 
While the present study offers valuable insights into the influence of camera perspectives on user experience and performance during immersive UAV piloting, several limitations should be acknowledged. First, the research was conducted with a moderate-sized sample in a controlled laboratory environment. As such, the generalizability of the findings to larger populations or to real-world, outdoor contexts with greater environmental complexity remains to be established. It should be noted, however, that the laboratory environment was deliberately selected because it provided the most straightforward context in which to create a high-fidelity digital twin. This ensured a reliable synchronization between the physical drone and its virtual counterpart, which was a prerequisite for studying perspective effects under controlled conditions. Future research should extend this work by replicating the study in more complex digital twins (e.g., urban outdoor spaces or dynamic environments) to assess how far the observed perspective effects generalize across different application domains.
Moreover, although a range of subjective and objective metrics were employed, some of the observed differences between camera perspectives did not reach statistical significance, often reflecting trends rather than robust effects. This suggests that future studies with larger sample sizes or more targeted participant selection may be necessary to detect subtler differences.
\color{black}
Another limitation concerns the relatively basic set of performance metrics used. While they provided an overview of task success and trajectory accuracy, they may not capture the full spectrum of user experience, cognitive load, or physical responses during drone operation.  Integrating physiological or behavioral measures—such as gaze tracking, heart rate, or real-time workload monitoring—could provide a richer and more nuanced understanding of user states.
A further limitation concerns the lack of systematic measurement of end-to-end latency, including pose-to-photon delays in the VR rendering pipeline and controller-to-drone transmission times. While manufacturer specifications and the observed stability of the setup suggest that delays remained within acceptable thresholds (<50~ms), we cannot exclude that latency may have affected user experience, particularly in demanding maneuvers. Future studies should incorporate precise latency measurement tools to quantify these delays and assess their impact on both subjective and objective outcomes.
\color{black}
\section{Conclusion}
 
This study explored how virtual camera perspectives affect user experience metrics, and performance during UAV piloting in an immersive digital twin. While third-person views tended to reduce workload and improve path accuracy, and first-person views enhanced immersion and smoothness of control, most effects were modest and not universally significant. However, our results converge with and extend a growing body of research indicating that the choice of camera viewpoint can critically shape both the subjective and objective aspects of drone operation~\cite{hing2010indoor,drury2006comparing}.
\color{black}
The observed variability in user preferences and performance highlights the importance of adaptable and user-centered interfaces, particularly as digital twin systems are adopted for complex, real-world UAV operations. While certain perspectives provided statistically significant advantages,  our findings highlight the importance of flexible, user-centered interfaces rather than the use of a single optimal perspective.
\color{black}

\begin{acks}
This work was supported by the European Union's Horizon Europe programme under grant number 101092875 ``DIDYMOS-XR'' (https://www.didymos-xr.eu) and from the Federal Ministry of Education and Research (BMBF) within the funding program “Innovative University” project „Transfer of digitalization expertise to the South Westphalia region“, virtual institute „Augmented and Virtual Reality“. The authors acknowledge the use of AI-based assistance (ChatGPT-5, OpenAI) for grammar checking, style harmonization, and sentence-level syntax editing during manuscript preparation.
\end{acks}

\bibliographystyle{ACM-Reference-Format}
\bibliography{sample-base}

\end{document}